\documentclass[aps,prl,twocolumn,superscriptaddress,showpacs]{revtex4-1}

\usepackage{amsmath,amssymb,bm}
\usepackage{graphicx,xcolor}

\newcommand*{\al}{\alpha}            
\newcommand*{\dl}{\delta}            
\newcommand*{\ka}{\kappa}            

\newcommand*{\DD}{\mathbf{D}}        
\newcommand*{\one}{\mathbf{1}}       
\newcommand*{\ox}{\otimes}           
\newcommand*{\x}{\times}             
\newcommand*{\7}{\dagger}            

\newcommand*{\ket}[1]{|#1\rangle}    
\newcommand*{\word}[1]{\quad\text{#1}\quad} 

\begin{document}

\title{Why doubly excited determinants govern configuration interaction\\
calculations of electron correlations}

\author{Carlos L. Benavides-Riveros}
\affiliation{Departamento de F\'isica Te\'orica, Universidad de
Zaragoza, 50009 Zaragoza, Spain}
\affiliation{Instituto de Biocomputaci\'on y F\'isica de
Sistemas Complejos, U. de Zaragoza, 50018 Zaragoza, Spain}
\affiliation{Physikalische und Theoretische Chemie,
Universit\"at des Saarlandes, 66123 Saarbr\"ucken, Germany}

\author{Jos\'e M. Gracia-Bond\'ia}
\affiliation{Departamento de F\'isica Te\'orica, Universidad de
Zaragoza, 50009 Zaragoza, Spain}
\affiliation{Instituto de Biocomputaci\'on y F\'isica de
Sistemas Complejos, U. de Zaragoza, 50018 Zaragoza, Spain}
\affiliation{Escuela de F\'isica, Universidad de Costa Rica,
11501 San Pedro de Montes de Oca, Costa Rica}

\author{Michael Springborg}
\affiliation{Physikalische und Theoretische Chemie,
Universit\"at des Saarlandes, 66123 Saarbr\"ucken, Germany}

\date{\today}

\begin{abstract}
Computational evidence shows that, when using natural orbitals to
study (dynamical and non-dynamical) electron correlation, determinants
with an odd number of excitations play a negligible role. Instead,
doubly excited determinants rule the rostrum in this kind of
configuration interaction calculations. We explain mathematically why
it must be so.
\end{abstract}



\maketitle  

\paragraph*{Introduction.}

Postulated by Pauli to explain the electronic structure of atoms and 
molecules, the \textit{exclusion principle} states that each quantum 
state cannot be occupied by more than one electron. As Dirac pointed
out, this principle emerges from antisymmetry imposed on the wave
function~\cite{Dirac}. The exclusion principle can be stated by saying
that the fermionic natural occupation numbers (NON), which are the
eigenvalues of the one-body reduced density matrix (arranged in
decreasing order $n_i \geq n_{i+1}$) must fulfill the constraint
$n_1 \leq 1$. In the sixties Coleman~\cite{Coleman} proved that this
inequality is necessary and sufficient for a one-body density matrix
to come from the reduction of an \textit{ensemble} $N$-body density
matrix, provided that $\sum_i n_i = N$.

In the Configuration Interaction (CI) picture, the antisymmetry of the
wave function is ensured by writing it as a linear combination of all
possible configurations,
\begin{equation}
\ket{\Psi} = \sum c_i\,\ket{\bm{i}},
\label{eq:CI} 
\end{equation}
where $\ket{\cdot}$ denotes a Slater determinant, in a given
spin-orbital basis. In this paper we use the basis of natural orbitals
(NO), the eigenvectors of the one-body reduced density matrix.

In a seminal article, Borland and Dennis observed generalized Pauli
conditions (GPC) for the rank-six approximation of a
\textit{pure-state} three-electron system~\cite{losprecursores}. The
NON satisfy the constraints:
\begin{gather}
n_j + n_{7-j} \leq 1, \text{ where } j \in \{1,2,3\};
\notag \\
\word{and} n_4 \leq n_5 + n_6.
\label{eq:constraints} 
\end{gather}

Then the question of possible GPC lay dormant for many years. Only a
few years ago, a standardized approach to them by profound
group-representation methods was devised by
Klyachko~\cite{Alturulato}. This reveals a rich substructure in
fermion systems, recently exploited in theoretical chemistry,
entanglement theory and ferromagnetism \cite{CGS13, Sybilla,
ChacraMaza, sarasaleva1, ultimochelin, inprocess, Klyachko13}.

\medskip

The Klyachko algorithm produces sets of linear inequalities for the
$m$~NON of the pure state
$\ket\Psi \in \wedge^n \mathcal{H}_m$ of $n$~electrons arranged in
$m$ spin orbitals, similar to those of~\eqref{eq:constraints}. Namely,
\begin{equation}
D^\mu_{n,m}(\bm{n}) 
= \ka^\mu_0 + \ka^\mu_1 n_1 +\cdots+ \ka^\mu_m n_m \geq 0,
\label{eq:Borden} 
\end{equation}
with $\bm{n} = (n_1,\dots,n_m)$ and integer coefficients $\ka^\mu_j$.
For instance, the generalization $n_j + n_{2k+1-j} \leq 1$ of the
first equations in~\eqref{eq:constraints} holds in \textit{any} even
rank $m = 2k$. The inequalities define a \textit{convex polytope} of
allowed states in~$\mathbb{R}^m$.

By definition, a \textit{pinned} system saturates completely some of
the GPC. That is, for some~$\mu$ indexing the corresponding equalities
\eqref{eq:Borden}, the condition $D^\mu_{n,m}(\bm{n}) = 0$ holds, and then
the system lies on one of the faces of the polytope. For such there is
a \textit{selection rule} given in~\cite{malquerido}, involving the
terms in the decomposition~\eqref{eq:CI}. To wit, define the operator
\begin{align}
\DD^\mu_{n,m} 
= \ka^\mu_0 \one + \ka^\mu_1 a^\7_1 a_1 +\cdots+ \ka^\mu_m a^\7_m a_m,
\label{eq:Oper} 
\end{align}
where $a^\7_i$ and $a_i$ are the creation and annihilation fermionic
operators for the state~$i$. Given a system satisfying
$D^\mu_{n,m}(\bm{n}) = 0$, each Slater determinant in the
expansion~\eqref{eq:CI} is an eigenfunction of $\DD^\mu_{n,m}$ with
eigenvalue~zero (say, effective configurations). In other words:
\begin{gather*}
\text{if } \DD^\mu_{n,m} \ket{\bm{i}} \neq 0, \text{ then } c_i = 0,
\\
\text{and therefore } \DD^\mu_{n,m} \ket{\Psi} = 0.
\end{gather*}
This extremely plausible statement, valid for nondegenerate NON,
actually needs proof, which is forthcoming~\cite{Grossetlopes}. It
enables the wave function to be described by \textit{Ans\"atze} that
drastically reduce the number of Slater determinants in the
CI~expansion. Numerical investigations for real atoms and
molecules~\cite{Sybilla, ChacraMaza} have already confirmed that
pinning often takes place. Even more mysteriously, there is a
remarkable prevalence of quasipinning (almost saturation of the
Klyachko inequalities) ---and not only for ground states.

\pagebreak

Now, recent evidence shows that,~when using the basis of NO (as
distinct from Hartree--Fock molecular orbitals, say) to study bond
weakening and breaking, doubly excited determinants are
\textit{dramatically enhanced} with respect to singly and triply
excited ones \cite{gritseslafilosofia}. In turn, this motivates the
introduction of a (quite successful) ``extended L\"owdin--Shull''
1-RDM functionals, sharing some of the simplicity of the original
L\"owdin--Shull formula for the wave function of a two-electron
system~\cite{LS1}. In this paper we argue that such an outstanding
phenomenon stems from Klyachko pinning, which eliminates first and
foremost oddly-excited configurations.

\medskip

\paragraph*{Rank-six three-electron systems.} For simplicity, we
consider three-electron systems, mostly described with the help of
restricted spin orbital bases. Besides the lithium isoelectronic
series, already thoroughly examined in~\cite{Sybilla,ChacraMaza}, we
base ourselves on data for perturbed lithium with broken spherical
symmetry, and for the dimer ion $\mathrm{He}_2^+$, for several
ranks~\cite{inprocess}.

Before proving our main result, it is instructive to understand it
first in the context of the Borland--Dennis case, with rank $m = 6$.
The general configuration possesses $\tbinom{6}{3}$ Slater
determinants. Now, due to $\sum_{i=1}^6 n_i = 3$, the Klyachko
inequalities
\[
n_1 + n_6 \leq 1, \quad 
n_2 + n_5 \leq 1, \quad
n_3 + n_4 \leq 1
\]
are in fact pinned. This selects a combination of \textit{eight}
states, living in $\mathcal{H}_2^{\ox 3}$. We may denote the three
pairs of NO by $\{\al_1,\al_6\}$, $\{\al_2,\al_5\}$, $\{\al_3,\al_4\}$. 
Note the following: if we decide that the spin of
the ground state is~$\uparrow$ (say), then with our basis we construct
nine corresponding eigenfunctions of~$S_z$; however, one of these
configurations belongs in the representation with $j = 3/2$. So we
have automatically obtained the correct counting of states. There are
three singly-excited determinants and a triply-excited one.

Furthermore, for the chosen basis of restricted spin orbitals, we are
able to prove~\cite{inprocess} that pinning of the last constraint
$n_4 = n_5 + n_6$ or $1 + n_3 = n_1 + n_2$ in~\eqref{eq:constraints}
\textit{always} applies. With that pinning, the three single
excitations, the triple excitation and one double excitation
disappear, so one needs only \textit{three} configurations, instead of
eight. Indeed: the operator $\one - a^\7_1 a_1 - a^\7_2 a_2 + a^\7_3
a_3$ does not ``kill'' singly-excited configurations, so these cannot
enter the wave function. It does not kill the triple configuration
either; nor does it kill the doubly-excited configuration
$\ket{\al_3\al_5\al_6}$. Thus only the configurations
$\ket{\al_1\al_4\al_5}$ and $\ket{\al_2\al_4\al_6}$, besides
$\ket{\al_1\al_2\al_3}$, are available.

In conclusion, one rather efficiently has, for any state constructed
according to our specification:
\[
\ket{\Psi}_{3,6} = a\,\ket{\al_1\al_2\al_3}
+ b\,\ket{\al_1\al_4\al_5} + c\,\ket{\al_2\al_4\al_6},
\]
with
\begin{align}
n_1 = |a|^2 + |b|^2 \geq n_2 = |a|^2 + |c|^2, &\quad n_5 = |b|^2,
\notag \\
n_3 = |a|^2 \geq n_4 = |b|^2 + |c|^2, &\quad n_6 = |c|^2,
\label{eq:penapenitapena} 
\end{align}
for the NON. (In the simpler, somewhat degenerate case $m = 5$, two
single excitations and, as before, one double excitation are
ineffective.)

\medskip

\paragraph*{Rank seven.} For higher ranks, we look at the theoretical
and computational situations in parallel. For $m = 7$ the Klyachko
setup of inequalities is still mercifully small. To wit, there are
only four constraints in $\wedge^3 \mathcal{H}_7$:
\begin{align*} 
D^1_{3,7} &:= 2 - n_1 - n_2 - n_4 - n_7 \geq 0, \\
D^2_{3,7} &:= 2 - n_1 - n_2 - n_5 - n_6 \geq 0, \\
D^3_{3,7} &:= 2 - n_1 - n_3 - n_4 - n_5 \geq 0, \\
D^4_{3,7} &:= 2 - n_2 - n_3 - n_4 - n_6 \geq 0.
\end{align*}
Notice that the Klyachko restrictions are \textit{consistent}, in that
lower-rank ones can be derived from higher-rank ones. For instance, if
$n_7 = 0$ above, then summing the second and third we obtain
$n_1 + n_6 \leq 1$; the second and fourth yield $n_2 + n_5 \leq 1$,
and so on: we recover all the Borland--Dennis relations for
$\wedge^3 \mathcal{H}_6$. To be sure, the original Pauli principle
$n_1 \leq 1$ follows from them, too.

Numerical investigations for the lithium series have shown the system
to be always pinned to the first constraint~\cite{Sybilla,ChacraMaza}.
Moreover, the second constraint happens to be nearly saturated.
Remarkably, for $\mathrm{He}_2^+$ in its ground state the situation
appears to be reversed: the second constraint is saturated exactly,
while the first one is saturated to a good
approximation~\cite{inprocess}.

\begin{table}[!b]
\centering      
{\scriptsize 
\begin{tabular}{c@{\hspace{2pt}}cccccccc}  
Rank & $n_1$  & $ n_2$ & $n_3$ & $n_4 (-3)$ & $n_5 (-3)$ &
$n_6 (-4)$   & $n_7 (-4)$ & $n_8 (-5)$ \\ [0.5ex] 
\hline\hline           
6 & 0.9993 & 0.9938 & 0.9932 & 6.75 & 6.14 & 6.1 & $-$ & $-$ \\
7 & 0.9989 & 0.9938 & 0.9928 & 6.56 & 6.38 & 7.6 & 5.7 & $-$ \\
8 & 0.9990 & 0.9953 & 0.9943 & 5.06 & 5.02 & 5.9 & 5.2 & 2.1 \\[1ex]
\hline 
\end{tabular} }
\caption{NON corresponding to rank-six up to rank-eight approximations 
for $\mathrm{He}_2^+$ at its equilibrium geometry.}
\label{table:He78} 
\end{table}

Table \ref{table:He78} contains the NON for rank-six up to rank-eight
approximations for the dimer ion $\mathrm{He}_2^+$, as computed using
a STO-6G basis set~\cite{inprocess}. For rank-seven, the second GPC is
completely saturated, while the first GPC belongs to a highly
saturated regime:
$$
D^1_{3,7} = 2.41 \x 10^{-6}. 
$$
The other two constraints belong to a different scale of quasipinning:
we had already detected these ``successive scales of quasipinning''
in~\cite{Sybilla}. In fact,
$$
D^3_{3,7} = 3.18 \x 10^{-4} \word{and} 
D^4_{3,7} = 8.24 \x 10^{-4}.
$$

It seems fair to conclude that there is a tendency to strong
quasipinning of the two first GPC in this approximation. If both were
saturated, then $1 + n_3 = n_1 + n_2$ would follow. Indeed, we would
have:
\[
2 = n_1 + n_2 + n_4 + n_7 \word{and}
2 = n_1 + n_2 + n_5 + n_6.
\]
Summing these two equalities, we see that
\[
4 = 2n_1 + 2n_2 + n_4 + n_5 + n_6 + n_7 = 3 - n_3 + n_1 + n_2,
\]
where we have used $\sum_i n_i = 3$. This means again that all singly
excited and triply excited determinants are suppressed, and the number
of effective configurations, all of which are doubly excited, drops
sharply.

A telling example of further disappearance of excitations in the
context of $m = 7$ is discussed by Klyachko in~\cite{malquerido}. It
happens in the first excited state of~$\mathrm{Be}$ with spin data
$(\mathbf{S},S_z) = (1,1)$, whose first occupation number is frozen to
one, so we may regard it as a three-electron system. Numerical
calculations for this state suggest further pinning. Imposing
saturation of the third remaining constraint, four configurations are
left:
\begin{align} 
n_1 = |a|^2 + |b|^2 +  |d|^2 &\geq n_2 = |a|^2 + |c|^2, 
\label{eq:Gordiano} 
\\
n_3 = |a|^2 &\geq n_4 = |b|^2 + |c|^2, 
\notag \\
n_5 = |b|^2 &\geq n_6 = |c|^2 + |d|^2, \quad  n_7 = |d|^2,
\notag 
\end{align}
with $\ket{\Psi}_{3,7}$ being 
\[
a\,\ket{\al_1\al_2\al_3} + b\,\ket{\al_1\al_4\al_5}
+ c\,\ket{\al_2\al_4\al_6} + d\,\ket{\al_1\al_6\al_7}. 
\]
The case $\DD^4_{3,7}\ket\Psi=0$, when both $D^1_{3,7}$ and
$D^2_{3,7}$ are saturated, would be similar, with
$\ket{\al_2\al_5\al_7}$ replacing $\ket{\al_1\al_6\al_7}$.
Formulas~\eqref{eq:penapenitapena} come smoothly
from~\eqref{eq:Gordiano} when $d = 0$.

This is perhaps the place to invoke evidence from the toy model of
spinless ``fermions'' on the line, subjected to a harmonic potential
and a harmonic interaction, studied in~\cite{CGS13}. The strength of
the latter interaction can be described by a suitable parameter~$\dl$,
whose vanishing implies that the ground state is a single determinant
with trivial NON. Perturbatively in the interaction, it is found for
$N = 3$ that corrections to the NON are of order $\dl^4$. However, the
relation $1 + n_3 = n_1 + n_2$ is violated only at order~$\dl^8$. So
there also quasipinning is patent; on the other hand, due to the
existence of spin, real electronic systems are more rigidly pinned.

\medskip

\paragraph*{Rank eight.}
Let us go now to $m = 8$. The dimension of the Hilbert space 
is $\tbinom{8}{3}$. Putting aside again the issue of spin
contamination, the sector in which we are interested contains
$24$~configurations, corresponding to
$\wedge^2 \mathcal{H}_4 \ox \mathcal{H}_4$, of which clearly $7$~are
singly excited and $3$ are triply excited. The condition
$1 + n_3 \geq n_1 + n_2$ still holds.

Table~\ref{table:M8} contains the numerical values of the first ten
(of~$31$) GPC for the rank-eight approximation to the ground state of
the molecule~$\mathrm{He}_2^+$ and lithium \cite{Sybilla, inprocess}.
The constraint $D^2_{3,8} \geq 0$ appears to be saturated exactly for
the diatomic ion; and the constraints
\begin{align*}
D^1_{3,8} \geq 0, \quad D^5_{3,8} := 1 - n_1 - n_2 + n_3 \geq 0,
\end{align*}
nearly so. For the lithium isoelectronic series, we have chosen to
show data obtained by working with unrestricted spin orbitals.
Restricted ones actually yield better values for the energy and
exhibit pinning. The point is that even by working with unrestricted
ones, the same constraints are also very nearly saturated.

\begin{table}[!t]
\centering    
{\normalsize 
\begin{tabular}{l@{\hspace{2em}}c@{\hspace{2em}}c}  
 \qquad  \qquad GPC  & $\mathrm{He}_2^+$   & Li \\ [0.5ex] 
\hline\hline 
$D^1_{3,8}    = 2 - n_1 - n_2 - n_4 - n_7$ & 0.0259 & 0.0017 \\
$D^2_{3,8}    = 2 - n_1 - n_2 - n_5 - n_6$ & 0.0000 & 0.0200 \\
$D^3_{3,8}    = 2 - n_2 - n_3 - n_4 - n_5$ & 0.1793 & 0.0671 \\
$D^4_{3,8}    = 2 - n_1 - n_3 - n_4 - n_6$ & 0.9036 & 0.0894 \\
$D^5_{3,8}    = 1 - n_1 - n_2 + n_3$       & 0.0048 & 0.0200 \\
$D^6_{3,8}    = 1 - n_2 - n_5 + n_7$       & 0.1582 & 0.0854 \\
$D^7_{3,8}    = 1 - n_1 - n_6 + n_7$       & 0.8826 & 0.1078 \\
$D^8_{3,8}    = 1 - n_2 - n_4 + n_6$       & 0.1841 & 0.0671 \\
$D^9_{3,8}    = 1 - n_1 - n_4 + n_5$       & 0.9084 & 0.0894 \\
$D^{10}_{3,8} = 1 - n_3 - n_4 + n_7$       & 1.0619 & 0.1548 \\[1ex]
\hline    
\end{tabular} }
\caption{First ten GPC (${}\x 10^3$) for $\wedge^3 \mathcal{H}_8$ and
the observed numerical values for $\mathrm{He}_2^+$ \cite{inprocess}
and lithium \cite{Sybilla}.}
\label{table:M8} 
\end{table}

The ``unreasonable effectiveness'' of the single quasipinning
$1 + n_3 \simeq n_1 + n_2$ is here again enough to suppress the odd
excitations, obtaining a reduction to $13$ (the strongly occupied one
plus $12$ doubly excited) configurations. The operator
$$
\DD^5_{3,8}= \one - a^\7_1 a_1 - a^\7_2 a_2 + a^\7_3 a_3
$$ 
does kill $12$ double excitations:
{\small
\begin{alignat*}{2}
\ket{\al_1\al_4\al_5}, &\quad 
\ket{\al_1\al_4\al_6}, &\quad
\ket{\al_1\al_4\al_8}, &\quad
\ket{\al_1\al_7\al_5},
\\
\ket{\al_1\al_7\al_6}, &\quad
\ket{\al_1\al_7\al_8}, &\quad 
\ket{\al_2\al_4\al_5}, &\quad
\ket{\al_2\al_4\al_6},
\\
\ket{\al_2\al_4\al_8}, &\quad 
\ket{\al_2\al_7\al_5}, &\quad 
\ket{\al_2\al_7\al_6}, &\quad
\ket{\al_2\al_7\al_8},
\end{alignat*}}%
which are the survivors. The double excitation $\ket{\al_3\al_4\al_7}$
drops out, as well. Further (quasi)pinning selects out other double
excitations; we refer to~\cite{inprocess} for that.

\begin{figure}[!t]
\centering 
\includegraphics[width=9cm]{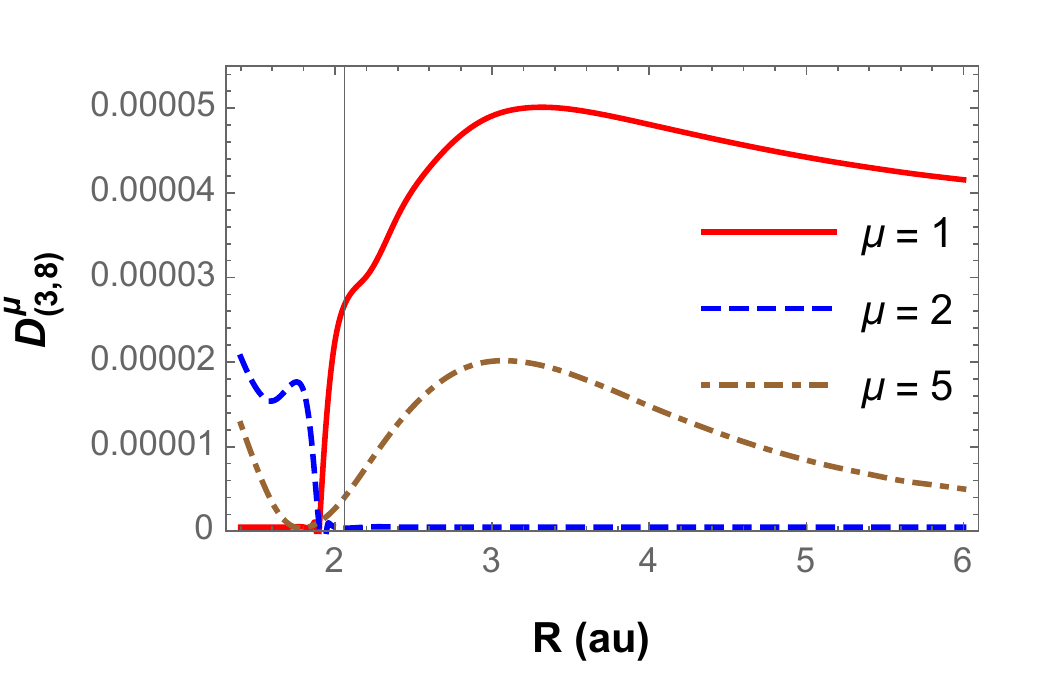}
\caption{(Color online)
$D^1_{3,8}$, $D^2_{3,8}$ and $D^5_{3,8}$ for $\mathrm{He}_2^+$ as
functions of the interatomic distance (atomic units are used).
The vertical line marks the equilibrium bond length.}
\label{figurerank81} 
\end{figure}

Figure \ref{figurerank81} exhibits the behavior of those three GPC as
functions of the $\mathrm{He}_2^+$ bond length. Notice the sudden,
intriguing crossover of two constraints at lengths smaller than that
of equilibrium (namely, $2.06$~au). This apparent quenching of degrees
of freedom deserves further investigation.


We summarize our findings in a quite parsimonious proposition.

\begin{quote}
\itshape The wave function of a three-fermion system, whose NON
satisfy the saturated Borland--Dennis--Klyachko condition $1 + n_3 =
n_1 + n_2$, contains no odd excitations.
\end{quote}

\paragraph*{Proof.}
Let us write the wave function as follows,
with $1 \leq i < j < k \leq m$ always:
\[
\ket{\Psi} = \sum_{i<j< k} c_{ijk}\,\ket{\al_i\al_j\al_k}; 
\word{so} n_a = \sum_{a \in\{i,j,k\}} |c_{ijk}|^2.
\]
Then $1 + n_3 = n_1 + n_2$ implies 
\begin{align*}
& \sum_{i<j<k} |c_{ijk}|^2 + \sum_{3 \in\{i,j,k\}} |c_{ijk}|^2
\\
&\qquad = \sum_{1 \in\{i,j,k\}} |c_{ijk}|^2 
+ \sum_{2 \in\{i,j,k\}} |c_{ijk}|^2.
\end{align*}
On the left hand side the amplitudes of the Slater determinants
containing the third natural orbital appear twice. In order to reach
this condition on the right, those amplitudes must correspond to
Slater determinants containing both the first and the second natural
orbitals. The only determinant doing that is $\ket{\al_1\al_2\al_3}$.

The remaining amplitudes on the left do not contain~$\al_3$. More
importantly, they appear only once. This implies that amplitudes
corresponding to Slater determinants of the type
$\ket{\al_1\al_2\al_x}_{x \neq 3}$ do not appear in the CI wave
function. Then the wave function, besides $\ket{\al_1\al_2\al_3}$,
only contains double excitations of this state, fixing moreover either
$\al_1$ or~$\al_2$.

\begin{widetext}
The wave function, subject to the condition $1 + n_3 = n_1 + n_2$,
then reads:
\[
\ket{\Psi}_{3,m} = c_{123}\,\ket{\al_1\al_2\al_3}
+ \sum_{4 \leq j < k \leq m} [c_{1jk}\, a^\7_j a_2 a^\7_k a_3
+ c_{j2k}\, a^\7_j a_1 a^\7_k a_3] \,\ket{\al_1\al_2\al_3}.
\]
Of course in practice we will not have $1 + n_3 = n_1 + n_2$ exactly
most of the time; but all the evidence so far available points to very
strong quasipinning here.
\end{widetext}

\paragraph*{Conclusion.} For four-electron molecules, the GPC
$2 + n_4 \geq n_1 + n_2 + n_3$ holds. If it is nearly saturated, as
in the case of the excited state of $\mathrm{Be}$ already discussed,
an almost identical argument to the above shows that simply, triply
and quadruply excited configurations are suppressed~\cite{inprocess}.
Many even-number electron systems fulfill the Smith identities
$n_1 = n_2$, $n_3 = n_4$, \dots. Quasipinning of the last indicated
GPC in this case translates simply into $n_1\simeq1$. The tug-of-war
between energy minimization and Pauli kinematics often means that some
electrons are frozen in lower shells and active spaces of smaller
dimension emerge~\cite{ChacraMaza,ultimochelin}.
Then the ``precipitous drop'' in single excitations seen in the
analysis of $\mathrm{BH}$~\cite{gritseslafilosofia}, with two
electrons frozen, leaves little doubt that the mechanism just
described is at~work there. Molecules with higher number of electrons
and multiple bonds constitute the next frontier, already being 
explored.

\begin{acknowledgments}
CLBR was supported by Colombian Department of Sciences and Technology.
He very much appreciates the warm atmosphere of the Physikalische und
Theoretische Chemie group at Saarlandes Universit\"at. JMGB was
supported by the Universidad de Costa Rica through its von~Humboldt
chair. The authors are most grateful to David Gross, Stephan Kohaut
and Joseph C.~V\'arilly for helpful comments and illuminating
discussions.
\end{acknowledgments}

\end{document}